\documentclass[useAMS,usenatbib]{mn2e}
\usepackage{aas_macros}
\usepackage{graphicx}
\newcommand{\carma}{{\small CARMA}}
\newcommand{\atca}{{\small ATCA}}
\newcommand{\vla}{{\small JVLA}}
\newcommand{\alma}{{\small ALMA}}
\newcommand{\swift}{{\small \it Swift}}
\newcommand{\bat}{{\small {\it Swift}/BAT}}

\title[Swift/BAT AGN at 100\,GHz]
{The mm-wave compact component of AGN}
\author[E. Behar et al.]
{Ehud Behar$^{1}$\thanks{E-mail: behar@physics.technion.ac.il}, Stuart Vogel$^{2}$,
Ranieri D. Baldi$^{3}$, Krista L. Smith$^{4}$, 
\newauthor and Richard F. Mushotzky$^{2}$
\\
$^{1}$Department of Physics, Technion, Haifa, Israel\\
$^{2}$Department of Astronomy, University of Maryland, College Park, USA\\
$^{3}$School of Physics and Astronomy, University of Southampton, UK\\
$^{4}$Einstein Fellow, Kavli Institute for Particle Astrophysics and Cosmology, Stanford University, USA
}
\begin{document}


\pagerange{\pageref{firstpage}--\pageref{lastpage}} \pubyear{2014}

\maketitle

\label{firstpage}

\begin{abstract}
mm-wave emission from  Active Galactic Nuclei (AGN) may hold the key to understanding  the physical origin of their radio cores. 
The correlation between radio/mm and X-ray luminosity may suggest a similar physical origin of the two sources.  
Since synchrotron self absorption decreases with frequency, mm-waves probe smaller length scales than cm-waves.
We report on 100\,GHz (3\,mm) observations with \carma\ of 26 AGNs selected from the hard X-ray \bat\ survey.
20/26 targets were detected at 100\,GHz down to the 1\,mJy (3$\sigma$) sensitivity, which corresponds to optically thick synchrotron source sizes of $10^{-4} - 10^{-3}$ pc.
Most sources show a 100\,GHz flux excess with respect to the spectral slope extrapolated from low frequencies.
This mm spectral component likely originates from smaller scales than the few-GHz emission.
The measured mm sources lie roughly around the $L_{\rm mm}$ (100 GHz) $\sim 10^{-4}L_X$ (2--10~keV) relation, similar to a few previously published X-ray selected sources, and hinting perhaps at a common coronal origin. 
\end{abstract}


\section{Introduction}
\label{Introduction}

Radio loud (RL) active galactic nuclei (AGN) are known for their relativistic jets that often extend beyond the nucleus.  
Radio emission from Radio Quiet (RQ) AGN \citep[as defined by][]{kellerman89} is lower by several orders of magnitude than that of RL AGN.  
RQ AGN show, on average, smaller radio structures some of which are often unresolved down to
parsec scales \citep{blundell96, blundell98, ulvestad01, lal04, nagar99, nagar02, anderson04, ulvestad05_n4151, singh13, kharb15},
and only sub-relativistic velocities \citep{middelberg04, ulvestad05_rqq}.  
The physical origin of the radio emission in RQ AGN, whether a downscaled version of the RL, collimated jets \citep{barvainis96, gallimore06}, 
or coronal emission from magnetic activity above the accretion disk \citep{Field93}, remains to be resolved.
In some sources, a combination of several spectral components may be present around the nucleus \citep{barvainis96, gallimore04, giroletti09, panessa13, smith16}.  
For the sake of clarity of discussion, we distinguish emission from a jet that is well-collimated (and likely relativistic), 
from coronal emission that arises from hot ($\sim 10^9$K) gas, but could have an outflow that is not well-collimated (nor relativistic).

A connection between the radio and X-ray emission in RQ AGN is
suggested by the correlation of the radio luminosity $L_R$ at 5 GHz and the X-ray luminosity $L_X$  
\citep{brinkmann00, salvato04, wang06, panessa07, panessa13}.
On scales of a few mas (sub-pc) in nearby Seyferts, the core radio to X-ray correlation is not obvious \citep{panessa13}, although the high-resolution studies still suffer from low-number statistics.
\citet{laor08} used the PG quasar sample \citep{schmidt83} to demonstrate that $L_R (\equiv \nu L_\nu$ at 5~GHz) and
$L_{\rm X}$ (0.2--20~keV) are correlated over a large range of AGN luminosity, with a scatter of about an order of magnitude. 
More importantly, the correlation follows the well established relation for coronally
active cool stars $L_{\rm R}/L_{\rm X}\sim 10^{-5}$ \citep{guedel93}. 
This suggests that radio emission from RQ AGN may be due to
magnetic coronal activity, akin to that of stellar coronae.  
The $L_R/L_X\sim 10^{-5}$ relation in cool-stars is accepted as a manifestation of
coronal heating by electrons energized by magnetic reconnection,
which subsequently gives rise to X-ray emission.
Thus, the correlation presented in \citet{laor08} over 20 orders of
magnitude in luminosity raises the possibility that radio emission in
RQ AGN may also be related to coronal, magnetic activity.
Theoretical study of the magnetic dynamos in these vastly different systems suggests
a common mechanism could be at work \citep{Blackman00}.

There are, however, profound phenomenological differences between X-ray and radio emission
of stellar coronae and those of AGN.
X-ray spectra of stellar coronae are thermal with $T \sim 10^6$~K \citep{Catura75}, 
while the non-thermal (Comptonized) spectra of AGN imply $T \sim10^9$~K \citep{Shapiro76, Lightman78}.  
Moreover, while X-rays from AGN vary dramatically over short time scales \citep{Tananbaum78}, 
the variability at 5 and at 8.5~GHz is much slower and smaller in amplitude \citep[e.g.,][]{Anderson05}.  
On the other hand, if radio emission from RQ AGN is due to self absorbed synchrotron, 
and since absorption decreases with frequency, higher frequencies would originate from smaller core regions and vary more.
The decreasing size with frequency, of an optically thick self-absorbed synchrotron source, can be written as:

\begin{equation}
R_\mathrm{pc} \simeq 0.54L^{1/2}_{39}\nu^{-7/4}_\mathrm{GHz}B^{1/4}
\label{eq:Rpc}
\end{equation}

\noindent where $R_\mathrm{pc}$ is the size in pc, $\nu_\mathrm{GHz}$ is the observed frequency in GHz,
$L_{39} = \nu _\mathrm{GHz}L_{\nu_\mathrm{GHz}} /10^{39}$\,erg\,s$^{-1}$, and $B$ is the magnetic field in Gauss \citep{laor08}.  

As opposed to 5 and 8.5 GHz, higher frequency observations of RQ AGN are scarce.
\citet{barvainis96} observed RQ AGN up to 15~GHz.
About half of the RQ AGN in that sample appeared to have flat or inverted spectral components, 
and a hint of variability, suggesting opaque synchrotron emission from the nucleus.
\citet{park13} reach similar conclusions based on observations at 22 and 45 GHz.
Above 300\,GHz (sub-mm, FIR) there is a steep rise of the spectrum due to the low frequency tail of thermal dust emission from the host galaxy \citep{barvainis92, hughes93}, whose tail around 100\,GHz is typically negligible even in low-luminosity AGN \citep{doi11, Wu18}, let alone in strong X-ray emitters \citep{doi16}.

The waveband between 30--300~GHz, crudely referred to here as the mm waveband, has remained largely unexplored.
Notable exceptions are \citet{Field93} and \citet{DiMatteo97}, who addressed
the inevitable cyclotron or synchrotron emission from the magnetically confined 
thermal and non-thermal electrons,
but did not specifically ascribe the radio emission of RQ AGN to coronal activity.
In the past few years, there have been increased efforts to study the mm-band of RQ AGNs.
\citet{doi05, doi11} observed a sample of low luminosity AGN and early type galaxies, both RQ and RL, some of which showed a 
flat or even inverted spectrum at high frequencies, indicating genuine optically thick core emission.  
For studying the coronal conjecture, however,  it is beneficial to select X-ray active (and highly variable) RQ AGN, where the X-ray emission is ascribed to a compact corona.

In an exploratory program with the \carma\ and \atca\ telescopes, 95\,GHz continuum emission was detected from eight X-ray selected RQ AGN at the few mJy level \citep{behar15}. 
For the most part, the 95\,GHz flux density exceeded that expected from interpolating the low-frequency steep slope by up to a factor of $\sim 7$. 
This was called the high-frequency excess.
It was interpreted as evidence for a compact, optically thick core that is superimposed on (at the few arcsecond resolution of \carma\ and \atca ) the steep spectrum of more extended structures that dominate at lower frequencies.
The size of the 95\,GHz sources were estimated from Eq.~\ref{eq:Rpc} (assuming for simplicity $B = 1$G) to be of less than a light day, and between 10 and 1000 gravitational radii, which is comparable to the X-ray source.
That work also found the sources to closely follow $L_R$ (95 GHz) $\sim 10^{-4}L_X$ (2--10~keV), reminiscent of the $10^{-5}$ ratio at 5\,GHz. 

A high-frequency excess at 115\,GHz was observed in the lenticular galaxy, likely hosting a low-luminosity AGN, NGC\,1277 \citep{scharwachter16}.
A high-frequency excess was also noted in the Seyfert galaxy NGC\,985 by \citet{doi16}, who found that the steep spectrum turns over and becomes inverted around 10\,GHz.
These authors discussed several physical models that could produce the observed excess, most of which were ruled out, e.g., dust emission, free-free emission from the corona or from the broad-line region, and thermal synchrotron from an accretion flow.
They conclude that a jet would need to be free-free absorbed, and that self-absorbed synchrotron from the corona is also possible.
An interesting emission excess at 500\,$\mu$m (600\,GHz), which is correlated with hard X-ray luminosity, was reported by \citet{Shimizu16} for \bat\ AGN with respect to normal star forming galaxies.
This excess suggests that the AGN emission is still significant into the FIR.
Theoretical works \citep{inoue14, raginski16} further highlight the potential of mm observations for detecting coronal magnetic activity on AGN accretion disks.

\citet{baldi15} found that the nearby Seyfert NGC\,7469 varies at 95\,GHz (confidence level 99.98\%) on a time scale of days, while being apparently steady for years at 5\,GHz (P\'{e}rez-Torres et al., 2009).
\alma\ observations of NGC\,7469 indicate that the high-frequency excess remains up to $\sim 350$\,GHz, and that it originates from a bright core at sub-arcsec scales \citep{izumi15}.
To date, the daily variability, which is also observed in X-rays, still provides the most stringent constraints on the unresolved core sizes. 

In this work, we report on 100\,GHz \carma\ (the Combined Array for Research in Millimeter-wave Astronomy) observations of \bat\ AGN. 
\citet{smith16} observed 65 \bat\ AGN with the \vla\ at 22\,GHz and detected 62 of them.
All sources were found there to have bright cores of less than 1\arcsec , and about half also had extended structures. 
The goal here is to measure 100\,GHz continuum luminosities, to calculate the implied source sizes, and to study the relation to the well measured \bat\ hard X-ray luminosities. 
The remainder of the paper is organized as follows:
In Sec.~\ref{sec:sample} we describe the sample and data reduction.
In Sec.~\ref{sec:results} we present the 100\,GHz source fluxes and their relation to those at lower frequencies and to the X-rays. 
The conclusions and outlook are given in Sec.~\ref{sec:discussion}.

\section{Sample and Observations}
\label{sec:sample}

The Burst Alert Telescope (BAT) on board \swift\ observes all AGN (types 1 and 2), but the most Compton thick ones, between  14--195~keV. 
The current sample is drawn from the 70-month \bat\ AGN survey \citep{baumgartner13}.
It therefore selects high hard-X-ray-flux AGN with moderately high black-hole masses ($>10^6M_\odot$) and accretion rates ($L/L_{\rm Edd} >10^{-3}$), 
See \citet{koss11b, koss17} for more details.
These sources presumably all have an active accretion disk and an X-ray corona.
The present sub-sample was selected to include the luminous X-ray sources ($\log L_{\rm BAT} > 42.5$ in erg\,s$^{-1}$), which were further selected to have a bright CO J = 3 -- 2 line (Koss et al., in preparation), 
so that \carma\ can measure their CO J = 1 -- 0 line at 115.271\,GHz (to be reported separately). 

No further down-selection was applied based on radio loudness, but the vast majority of the sources are RQ.
A complete survey of the radio loudness of \bat\ AGN, unfortunately, does not exist \citep[but see][]{Wong16}.
Using the NED database\footnote{The NASA/IPAC Extragalactic Database (NED) operated by the Jet Propulsion Laboratory, California Institute of Technology.}, we find three sources that seem to be highly RL:  2MASX\,J0759+2323, MCG\,--01-40-001, and NGC\,5106, and a few others that are moderately RL (e.g., NGC\,2110).
As shown below, this property is generally not borne out by their mm to X-ray luminosity ratios, which motivates us at this point to retain all sources, and to treat them on an equal footing regardless of radio loudness.
The observation log is given in Table~\ref{obslog}.
Source names and Seyfert types are taken from the \bat\ 70-month survey. 

\begin{table*}
 \centering
\caption{Observation log}
\begin{tabular}{lccc}
  \hline
Object   &  $z$ & Seyfert   & CARMA  \\
  ~~(1) & & type & obs. date \\
  \hline
2MASX J0444+2813  & 0.0114 &  2 &  2013-10-16  \\
2MASX J0759+2323 & 0.0292 & 2 &  2013-11-08\\
2MASX J1937-0613   & 0.0103 & 1.5 &  2013-10-16 \\
MCG --01-40-001   & 0.0227 &2 & 2013-11-09    \\
MCG +04-48-002   &  0.0139 &2 &  2013-10-14 \\
MCG +08-11-011   & 0.0205 &1.5 & 2012-12-12  \\
Mrk\,79 & 0.0223 & 1.2 &  2013-10-12 \\
Mrk\,520   & 0.0266 &  1.9 &  2013-10-16   \\
Mrk\,618   & 0.0355 & 1 &  2012-12-09  \\
Mrk\,732   &  0.0292&1.5 &   2012-12-11   \\
Mrk\,739    & 0.0299 &1 &  2013-10-27   \\
Mrk\,1210   & 0.0135 &2 &  2013-11-09  \\
NGC\,2110   & 0.0078 &  2 &  2013-10-12    \\
NGC\,2992   &  0.0077 &2 &   2013-11-13 \\
NGC\,3516   & 0.0088 & 1.5 & 2013-01-05  \\
NGC\,3786 & 0.0089 & 1.8 & 2013-11-16 \\
NGC\,4388   & 0.0084 & 2 &  2013-01-06 \\
NGC\,4593 & 0.0090 & 1 &  2013-11-18 \\
NGC\,5106   & 0.0319 & LINER  &  2013-01-06  \\
NGC\,5290    & 0.0086 & 2 & 2013-10-21  \\
NGC\,5506   & 0.0062 &1 &  2013-01-02   \\
NGC\,5610   & 0.0169 & 2 &  2013-10-26 \\
NGC\,5728 & 0.0093 & 2 & 2014-04-20 \\
NGC\,5995   & 0.0252 & 2 & 2012-12-30   \\
NGC\,7679   & 0.0171&  2 & 2013-10-18 \\
UGC\,07064  & 0.0250 &  1 & 2012-12-23 \\
  \hline
\end{tabular}
\label{obslog}
\begin{flushleft}
NOTE - (1) Name, redshift, and Seyfert type as in the \bat\ 58-month survey https://swift.gsfc.nasa.gov/results/bs58mon/ .
\end{flushleft}
\end{table*}

\carma , now decommissioned, was a 15 element interferometer consisting of nine 6.1~m antennas and six 10.4~m antennas, located in California (USA).
The \carma\ observations were performed in the C-array configuration, providing an angular resolution of $\sim 1\arcsec$ at 100\,GHz.
This beam corresponds in the present low-$z$ (median 0.016) sample to a physical size of less than a kpc (median of 350\,pc).

The MIRIAD software package \citep{sault95} was used for reduction of
the visibility data, including flagging data affected by shadowed
antennas, poor weather or antenna malfunctions.
After obtaining the phase and amplitude solution,
we apply them to the target source using standard procedures.
Eventually, we invert  and clean with natural weight the visibility data to obtain the map. 
The MIRIAD {\it imstat} task is used to measure the source point-source peak flux density, and its uncertainty.

Owing to the broad band observed, we are able to produce separate maps at $\sim 100$\,GHz and at $\sim 114$\,GHz.
We use here mostly the lower sideband, which has slightly better signal to noise ratio (S/N).
For a few sources, we are able to obtain the spectral slope between the two sidebands.

\section{RESULTS}
\label{sec:results}

The sources are generally unresolved. 
The average (and median) synthesized beam FWHM (Full Width Half Maximum) is $2\farcs 0 \times 1\farcs 2$ with a standard deviation (std) of $0\farcs 4 \times 0\farcs 14$.
We convolve the maps to 2\arcsec and 6\arcsec resolution to compensate for varying atmospheric seeing, and to get an idea of extended AGN emission, but find that the flux increases by only 9\% and 32\%, respectively, by median.
Gaussian fits with the {\it imfit} task also produce very similar results.
The median peak Gaussian flux is 2.2\,mJy compared to 2.1\,mJy for a point source ({\it imstat}), and the median de-convolved Gaussian FWHM is $1\farcs 6 \times 0\farcs 8$, also consistent with the aforementioned synthesized beams.
All evidence points to mostly unresolved cores, although there is a hint to some extended emission, as the total integrated Gaussian fits yield on average 50\% higher flux density than their peaks. 
NGC\,2110, which is known to have an extended radio jet \citep{ulvestad89}
has a Gaussian integrated flux of 17.8\,mJy, as compared to 10.8\,mJy peak flux, and 13\,mJy in the 2\arcsec extraction,
but it is the RL exception in this sample, rather than the rule.

\subsection{100\,GHz flux densities}
\label{100GHz}

Out of the 26 AGNs observed, 18 were detected at 100\,GHz, and upper limits of about 1\,mJy (3$\sigma$) were obtained
for the remaining 8.
Most of the sources have flux densities below 10\,mJy, with the median being 2.1\,mJy.
The highest flux of 24.1\,mJy is measured for NGC\,5106, which is  the (only) LINER in the sample (Table~\ref{obslog}),
and will be discussed in more detail in Sec.~\ref{sec:results}.

Since we are seeking emission from regions much smaller than the $\approx$1\arcsec resolution, we list the point-source peak flux densities in Table~\ref{flux}, but note that those are very similar to the convolved 2\arcsec and Gaussian fitted peak values, as described above.
We also include the 1$\sigma$ rms (root mean square) uncertainty that was measured away from the source in each map.
The average (and median) rms is 0.3\,mJy with a std of 0.1\,mJy.
We further estimate a $\sim 15 \%$ systematic flux uncertainty that needs to be added to the statistical rms.

\begin{table*}
\centering
\caption{mm-band and X-ray properties}
\begin{tabular}{lcccccccc}
\hline
Object& $F_{\rm 100\,GHz}$  &  $\log L_{\rm mm} $ & $R_{\rm pc}$ &  $F_{\rm 22\,GHz}$ & Spectral slope & Excess & $\log L_{\rm BAT}$ & $\log (L_{\rm mm} / L_{\rm X}$)\\
 & (mJy)  & (erg\,s$^{-1}$) & ($10^{-3\,}$pc) & (mJy) & & $\times$ & (erg\,s$^{-1}$) \\
& (1) & (2) & (3) & (4) & (5) & (6) & (7)  & (8)\\
  \hline
2MASX J0444+2813  &   2.12 $\pm$ 0.37  & 38.77$\pm$0.10 & 0.132 & 3.0  & $0.8\pm 0.1$ & 3.5 & 43.18 & --3.92 \\
2MASX J0759+2323 &  $< 0.75$ &  $<38.32$ & $<0.080$ & --- & --- & --- & 43.83 & $<-5.06$ \\
2MASX J1937-0613   &  1.52 $\pm$ 0.36  & 38.55$\pm$0.13 & 0.102 & 5.16 & $0.76\pm 0.1$ & 0.90 & 42.74 & --3.67\\
MCG --01-40-001   &  3.98 $\pm$ 0.49  & 39.63$\pm$0.09 & 0.361 & --- & $0.73 \pm 0.04$ & 0.42 & 43.58 & --3.48\\
MCG +04-48-002   &  $< 0.96$ & $<39.02$ & $<0.176$ & 0.44& $1.5 \pm 0.05$ & $<6.8$ & 43.53 & $< -4.04$\\
MCG +08-11-011  & 7.53 $\pm$ 0.19  & 39.82$\pm$0.07 & 0.448 & ---& $1.3 \pm 0.3\,(1.5 \pm 0.3)$ & 13\,(22) & 44.16 & --3.91\\
Mrk\,79 &  1.54 $\pm$ 0.21 & 39.22$\pm$0.09 & 0.221 & 1.45 & $0.4 \pm 0.2\,(0.6 \pm 0.3)$ & 2.7\,(4.2) & 43.72 & --4.04\\
Mrk\,520   &  1.36 $\pm$ 0.26  & 39.32$\pm$0.11  & 0.255 &  ---  & $1.58\pm 0.08\,(1.47 \pm 0.07)$ & 38\,(12) & 43.69 & --3.91  \\
Mrk\,618   & $< 1.2$  &  $<39.27$ & $< 0.242$ & ---  & $1.00\pm 0.09\,(0.96 \pm 0.07)$ & $< 1.5\,(1.2)$ & 43.72 & $< -4.00$\\
Mrk\,732   &  1.99 $\pm$ 0.22  & 39.54$\pm$0.08  & 0.329 & --- & $0.5 \pm 0.2$ & 0.51 & 43.39 & --3.37\\
Mrk\,739    &  $<0.69$  & $<39.08$  & $<0.194$ & 0.31 & $0.77 \pm 0.1$ & $<8.6$ & 43.43 & $< -3.87$\\
Mrk\,1210   &   3.85 $\pm$ 0.28 & 39.19$\pm$0.07 &  0.216 & --- & $0.62\pm 0.05$ & 0.50 & 43.35 & --3.68\\
NGC\,2110   &  9.97 $\pm$ 0.35  & 39.12$\pm$0.07 & 0.196 & 42.17 & $0.3\pm 0.2\,(0.5\pm 0.1)$ & 0.4\,(0.6) & 43.63 & --4.05\\
NGC\,2992   & 5.14 $\pm$ 0.41  &  38.83$\pm$0.07 & 0.139 & 12.49 & $0.35\pm 0.3$ & 1.6 & 42.55 & --3.19\\
NGC\,3516   &  1.84 $\pm$ 0.29 & 38.50$\pm$0.10 & 0.095 & 3.70 & $0.2\pm 0.1$ & 1.1 & 43.31 & --4.33\\
NGC\,3786 & $< 1.02$ & $<38.24$& $<0.071$ & 0.72 & $1.0\pm 0.02$ &  $<2.0$ & 42.50 & $< -3.73$\\
NGC\,4388   &  3.58 $\pm$ 0.34  & 38.74$\pm$0.08 & 0.127 & 3.26 & $0.37\pm 0.15$ & 2.5 & 43.64 & --4.43\\
NGC\,4593 & $<1.02$ & $<38.20$  & $<0.068$ & --- &  $0.09\pm 0.12\,(0.4\pm 0.27)$ & $<0.24\,(0.38)$ & 43.20& $< -4.51$\\
NGC\,5106   &  24.1 $\pm$ 0.39 &  40.70$\pm$0.07 & 1.253 &  --- &  $0.4\pm 0.2$ & 1.5 & 43.54 & --2.37\\
NGC\,5290    & 1.42 $\pm$ 0.35  & 38.37$\pm$0.13 & 0.082 & 6.66 & $0.35\pm 0.15\,(0.6\pm 0.3)$ & 0.6\,(0.1) & 42.50 & --3.60\\
NGC\,5506   & 14.2 $\pm$ 0.33 & 39.08$\pm$0.07 & 0.185 & 48.5 & $0.9\pm 0.1$ & 1.9 & 43.31 & --3.75\\
NGC\,5610   &  1.05 $\pm$ 0.32 &  38.82$\pm$0.15  & 0.141 & --- & $0.77\pm 0.05$ & 1.24 & 43.09 & --3.77\\
NGC\,5728 & 1.61 $\pm$ 0.50 & 38.50$\pm$0.16& 0.096 & 4.08 & $0.33 \pm 0.04$ &  0.85 & 43.23 & --4.25\\
NGC\,5995   &   1.06 $\pm$ 0.20 &  39.15$\pm$0.11 & 0.207 & --- & $1.40\pm 0.05$ & 14 & 43.80 & --4.20\\
NGC\,7679   &  $<1.05$ &  $<39.15$ & $< 0.204$ & 0.46 & $1.2\pm 0.4\,(1.5\pm 0.3)$ & $< 3.4\,(5.8)$ & 43.00 & $< -3.35$ \\
UGC\,07064   &  $< 0.87$ &  $<38.75$ & $< 0.131$ & 0.61  & $0.92\pm 0.04$ & $<2.0$ & 43.28 & $< -4.05$ \\
  \hline
\end{tabular}
\label{flux}
\begin{flushleft}
NOTE - (1) Present \carma\ measurement, with 1$\sigma$ rms uncertainties. Upper limits are 3$\sigma$.
(2) $L_{\rm mm}  = \nu L_\nu$ at 100\,GHz; uncertainties include 15\% systematics.
(3) $R_{\rm pc}$ from Eq.~\ref{eq:Rpc} (with $B = 1$G)  in mpc.
(4) Core 22\,GHz flux density, mostly from \citet{smith16}.
(5) Spectral index $\alpha$ ($F_\nu \propto \nu^{-\alpha}$) based on archival data up to 22\,GHz, as available, slope in parenthesis refers to measurements with large beams.
(6) Ratio of observed 100\,GHz flux density to that expected from extrapolating the low-frequency slope; parentheses refer to corresponding slope for larger beam size.
(7) \bat\ X-ray luminosity in 14--195~keV band.
(8) $L_{\rm X}$ as obtained from Eq.~\ref{BATtoX}.
\end{flushleft}
\end{table*}

The consequent mm-wave luminosities $L_{\rm mm} = \nu L_\nu$ at 100\,GHz, are listed in the table, using the luminosity distances of the sources.
In the table we also list $R_{\rm pc}$ from Eq.~\ref{eq:Rpc} in units of $10^{-3}$\,pc (mpc).
These values can be associated with $\sim 10\%$ uncertainty, and they represent the actual physical source size if it is truly an optically thick synchrotron source.
For this estimate we use $B = 1$G.
In equipartition of the magnetic field with the AGN photon field, the magnetic fields will be 1 --2 orders of magnitude larger  \citep[See Eq. 21 in][]{laor08}.
However, the $B^{1/4}$ dependence (Eq.~\ref{eq:Rpc}) makes $R_{\rm pc}$ fairly insensitive to $B$,f and adopting equipartition values would only slightly increase its estimates.
It can be seen in Table~\ref{flux} that all sources are of the order of  $R_{\rm pc} \sim 0.3$\,mpc or less, implying light crossing times (and perhaps variability) of less than 1 day.
These size estimates, which totally depend on the assumption of synchrotron self-absorption, indicate cores that are six orders of magnitude smaller than the \carma\ beam size ($\sim 350$\,pc).

In the fifth column of Table~\ref{flux} we list the 22\,GHz core flux density for 15 sources, out of the present sample, although 4 of those were undetected at 100\,GHz.
Most of the 22\,GHz fluxes were measured with the \vla\ in the B- ($0\farcs 3$) or C- ($1\arcsec$) array configuration by \citet{smith16}, and a few will be published soon.
The beam size of these observations is not much smaller than the present one at 100\,GHz, which makes the comparison between the two meaningful.
The flux density of most sources declines between 22\,GHz and 100\,GHz, but three sources feature a rather flat or even inverted ($F_{\rm 100\,GHz} > F_{\rm 22\,GHz}$) spectrum between 22 and 100\,GHz (2MASX\,J0444+2813, NGC\,4388, and Mrk\,79).
NGC\,5506 was observed in 2003 at 22\,GHz at a high resolution of $0\farcs 1$ by \citet{tarchi11}, who measured a core flux of 22\,mJy, about half the current value. We suspect this is due to the 10 times smaller beam, but variability over 13 years can not be ruled out.

\subsection{Emergence of a new spectral component}
\label{component}

Assuming a power-law flux density spectrum $F_\nu \propto \nu ^{-\alpha}$, and using all available archival \vla\ measurements up to 22\,GHz, we were able to measure the spectral slope $\alpha$ for all of the sources, except 2MASX\,J0759+2323 that has only one frequency measured.
The resulting slopes are listed in column (5) of Table~\ref{flux}.
We use preferably A-array fluxes, but when larger beams are available we list the resulting slopes in parenthesis.
The two slopes are always consistent with each other within the (quoted) uncertainties.
The following column (6) in Table~\ref{flux} lists the high-frequency excess of the current \carma\ measured 100\,GHz flux density,
 with respect to the extrapolation of the low-frequency power-law slope.
Excess values in parenthesis correspond to the slopes in parenthesis.

\begin{figure}
\vskip -0.3cm
\hskip -1.4cm
\includegraphics[scale=0.38,angle=180]{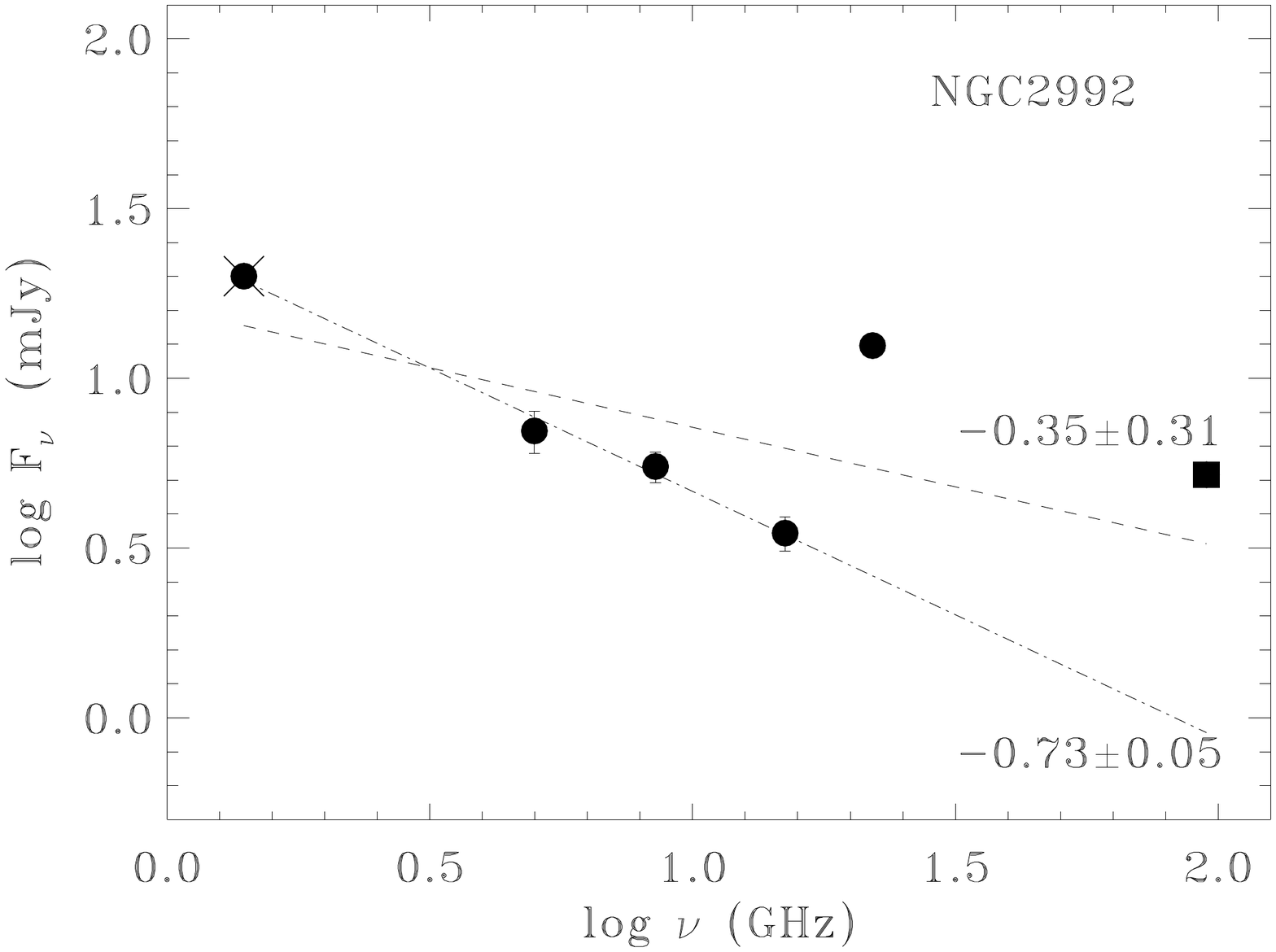} 
\vskip -0.7cm
\caption{Radio to mm spectrum of NGC\,2992 demonstrating the steep slope of $\alpha = 0.73$ at low frequencies and the high-frequency excess. The flux density at 1.4 and 5 GHz is from \citet{ulvestad84a}, 8.4\,GHz is from \citet{thean00}, 15\,GHz is from \citet{carral90}, and 22\,GHz is from \citet{smith16}.  The 100\,GHz measurement is from the present work. It can be seen that the spectrum already starts to rise at 22\,GHz, and that the slope fitted with this data point is much shallower ($\alpha = 0.35$). }
\label{ngc2992}
\end{figure}

We find that out of the 25 measured slopes, 12 detected sources (out of 18) show a high-frequency excess at 100\,GHz, and 6 (out of the 7) upper limits are consistent with such an excess.
Most sources have a 100\,GHz excess of a factor few over the extrapolated slope, similar to the results in \citet{behar15}, and indicating a strong mm spectral component. 
The illuminating example of NGC\,2992 is shown in Fig.~\ref{ngc2992}.
Its steep slope up to 16\,GHz turns over around the 22\,GHz measurement and the 100\,GHz flux density is above the extrapolated low-frequency slope by a factor of $\sim$6, and still by 1.6 if the 22\,GHz data point is included in the slope fitting.
Mrk\,520 and NGC\,5995 stand out with excess values of 38 and 14, respectively. 
These two sources are Seyfert 2's, and their high excess values are a result of their steep slopes of $\alpha \approx 1.5$ (see Table~\ref{flux}).

Mrk\,739 needs to be treated carefully, as it includes a binary AGN, separated by $\sim6\arcsec$ \citep{koss11a}.
Both west (W) and east (E) nuclei are resolved with the \vla\ and have steep spectral slopes of $\alpha \approx 0.8$; 
Mrk\,739W has a bright radio nucleus, and is slightly extended, which led \citet{koss11a} to associate it with starburst.
On the other hand, both nuclei are X-ray sources; 
Mrk\,739W has a hard obscured X-ray spectrum and varies within hours.
Both AGNs are faint at higher frequencies. 
At 22\,GHz, Mrk\,739E is detected at 0.31\,mJy while Mrk\,739W is not, the \vla\ 3$\sigma$ limit being 0.15\,mJy \citep{smith16}.
At 100\,GHz, there is a weak signal at Mrk\,739W, but it does not constitute a significant detection (see Table~\ref{flux}), which based on the steep slope is not expected (see the high excess value in the Table). 


It should be noted that the lower frequencies below 22\,GHz are often observed with different beam sizes, and over several decades,
which can make the notion of a spectral slope, and a high-frequency excess somewhat misleading.
A more conservative measure of the mm-wave component could be the 22\,GHz to 100\,GHz flux density ratio $F_{\rm 22\,GHz}/F_{\rm 100\,GHz}$, since these measurements were carried out within 1 - 2 years, and with similar beam sizes.
As seen in Table~\ref{flux}, three sources have a flat or inverted spectrum in this range. 
In fact, $F_{\rm 22\,GHz}/F_{\rm 100\,GHz}$ in the group of the ten sources that are detected at both frequencies has a median of 1.8, while spectral slopes of $\alpha = 1, 0.7$, and 0.5 would imply $F_{\rm 22\,GHz}/F_{\rm 100\,GHz} = 4.5, 2.9$, and 2.1, respectively.

For four sources, we were able to measure the spectral slope around 22\,GHz by exploiting the 8\,GHz \vla\ observed band, namely between 18-26\,GHz.
We find for 2MASX\,J0444+2813, Mrk\,79, NGC\,2110, and NGC\,3516 slopes of $-\alpha = +0.08, -0.67, -2.4$, and +0.27, respectively.
No meaningful uncertainties could be obtained for these slopes.
NGC\,2110 is an extended radio loud source. 
Indeed, its steep slope of --2.4 around 22\,GHz that continues to drop towards 100\,GHz (no excess in Table~\ref{flux}) is consistent with its radio/mm emission originating from an optically thin source (the jet).
The spectra of 2MASX\,J0444+2813 and NGC\,3516 have flattened out by 22\,GHz compared to the broad-band slopes (Table~\ref{flux}), while Mrk\,79 is still rather steep, but must turn over before 100\,GHz, where it has a significant excess (Table~\ref{flux}).

For eight sources, we are able to estimate the spectral slope ($-\alpha$) between the sidebands of the \carma\ data from 100 to 116~GHz.
All slopes are consistent with being approximately flat or even inverted, though some with great uncertainty.
The three best measured slopes with $\Delta \alpha < 0.5$ are MCG\,+08-11-011 ($-\alpha = -0.47 \pm\ 0.29$), NGC\,5106 ($-\alpha = -0.06 \pm\ 0.18$), and NGC\,5506 ($-\alpha = 0.56 \pm\ 0.28$, i.e. clearly inverted).
In summary, there is a general flattening of the spectrum at high frequencies.
Where exactly does the slope turn over remains to be seen, and may very well vary from one source to the other.
45\,GHz flux measurements with the \vla\  are essential to better characterize this emerging new mm-band spectral component, and frequencies above 100\,GHz are crucial for differentiating the AGN core emission from that of the dust in the FIR.
In the mean time, the bright mm-wave cores of RQ AGNs that are spectrally distinguished from their low-frequency radio emission  provide tentative evidence for a prevalent, compact, optically thick nucleus, which could be the manifestation of the accretion disk corona in mm waves, and a potential source of short-time variability.

\subsection{X-ray luminosities}

\begin{figure}
\includegraphics[scale=0.33,angle=-0]{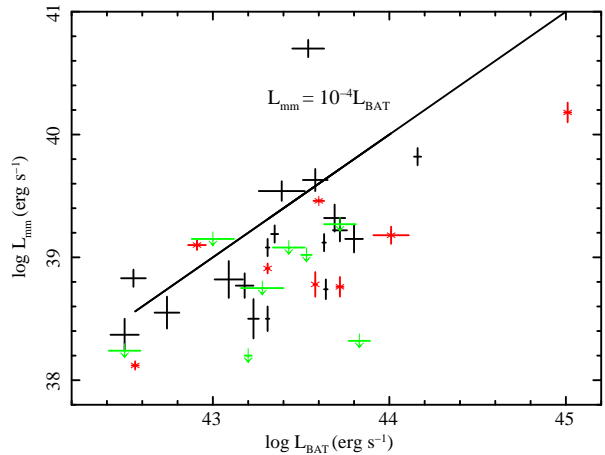} 
\caption{
$L_{\rm mm}  = \nu L_\nu$ at 100\,GHz vs. $L_{\rm BAT}$ (14--195~keV). Data include the sources from Tables \ref{flux} (black),
and \ref{previous} (red). Upper limits are marked with green down-pointing arrows. An $L_{\rm mm}  =  10^{-4}L_{\rm BAT}$ line is plotted for reference.
}
\label{LRLBAT}
\end{figure}

In this section, we compare the 100\,GHz luminosities to the X-ray ones, 
which are indicated in the last two columns of Table~\ref{flux}; 
uncertainties are not listed, but they are mostly small ($< 0.1$ dex).
A scatter plot of $L_{\rm mm} $ vs. $L_{\rm BAT}$ is presented in Fig.~\ref{LRLBAT},
which includes also seven AGNs that were measured previously at 95\,GHz \citep{behar15}, adding up to a total sample of 27 detections and 6 upper limits.  Upper limits are plotted as 3-sigma limits.
The previous paper includes eight AGNs (NGC\,5506 overlapping with the present sample) that were selected for their soft X-ray brightness and rapid variability, and are listed separately in Table~\ref{previous} below.
That selection naturally resulted in most of these being type 1 or type 1.5 Seyferts.
NGC\,5506 is classified in the \bat\ survey as a Seyfert 1.9, but is actually an optically obscured NLSy1 \citep{nagar02b}.
It is measured here at $14.2\pm 0.33$\,mJy, and previously at $10.0\pm 1.0$~mJy.
Ark\,564 from the previous sample is not in the \bat\ survey, but its hard X-ray luminosity $L_{\rm BAT}$ can be estimated to be $\log L_{\rm BAT}\approx 44.01$ (erg\,s$^{-1}$) from its soft X-ray 2 - 10~keV measured luminosity $L_{\rm X}$, using the empirical conversion formula of \citet{winter09}:

\begin{equation}
\log L_{\rm X} = 1.06 \log L_{\rm BAT} - 3.08
\label{BATtoX}
\end{equation}

\begin{table*}
\centering
\caption{Additional X-ray selected AGNs measured at 95\,GHz \citep[from][]{behar15}.}
\begin{tabular}{lcccccc}
\hline
Object& Seyfert & $F_{\rm 95\,GHz}$  &  $R_{\rm pc}$ & $\log L_{\rm mm} $ (95\,GHz) &  $\log L_{\rm BAT}$ & $\log (L_{\rm mm} / L_{\rm X}$)\\
 & type & (mJy)  & ($10^{-3}$pc)  & (erg\,s$^{-1}$) &  (erg\,s$^{-1}$)\\
  \hline
MR\,2251-178 & 1 & 1.6$\pm$0.3 &  0.70 & 40.18  &   45.01  &  --4.45  \\
NGC\,3783  & 1 & 3.1$\pm$0.7  & 0.15  & 38.78  &   43.58  & --4.33  \\
NGC\,5506   &  1.9 & 10$\pm$1.0 & 0.17  &  38.91 &   43.31 & --3.92  \\
NGC\,7469    & 1.2 & 4.95$\pm$0.16 & 0.32 & 39.46  &   43.60 & 3.68 \\
ARK\,564    & 1 & 1.14$\pm$0.19 & 0.23 & 39.18  &   44.01* & --4.39\\
NGC\,3227    &  1.5 & 4.1$\pm$0.24 & 0.07  & 38.12  &   42.56 & --3.91  \\
MRK\,766     &  1.5 & 1.98$\pm$0.17  & 0.16 &  39.10   &  43.91 & --3.30 \\
NGC\,5548    & 1.5 & 1.6$\pm$0.3 &  0.19 & 38.76  &   43.72 & --4.50 \\
  \hline
\end{tabular}
\label{previous}
\begin{flushleft}
* based on Eq.~\ref{BATtoX}.
\end{flushleft}
\end{table*}

The correlation between $L_{\rm mm} $ and $L_{\rm X}$ is interesting.
The values of the present work are listed in the last column of Table~\ref{flux}.
Using the hard X-ray band of \bat\ has the advantage of reducing the concern of absorption skewing an intrinsic correlation.
On the other hand, the luminosity range of the present sample is small, and it is far from being complete.
Hence, in Fig.~\ref{LRLBAT} we plot the $L_{\rm mm} = 10^{-4}L_{\rm BAT}$ line just to guide the eye.
The line appears to lie somewhat high with respect to the data, but there is much scatter.
In fact, a formal fit to the normalization (upper limits excluded) yields a somewhat weaker mean mm luminosity $L_{\rm mm}  = 10^{-4.19\pm 0.02}L_{\rm BAT}$ (90\% confidence uncertainties).
A fit of both the slope and the displacement in Fig.~\ref{LRLBAT}, yields a slope of $L_{\rm mm}  \sim L_{\rm BAT}^{0.92 \pm\ 0.04}$. 
A similar correlation, but between the 1.4\,GHz (on kpc scales) and 20--100~keV luminosities of a different hard X-ray selected AGN sample, yielded a slope of $1.2 \pm 0.15$ \citep[][Fig.~3 therein]{panessa15}, which is not very different from the present one. \citet{Wong16} find a correlation between $L_{\rm 1.4\,GHz}$ and $L_{\rm BAT}$, but do not report a direct slope.

\begin{figure}
\includegraphics[scale=0.33,angle=-0]{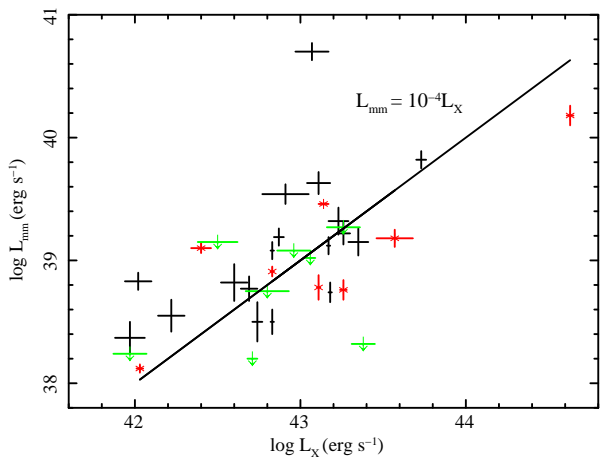} 
\caption{
$L_{\rm mm}  = \nu L_\nu $ at 100\,GHz vs. $L_{\rm X}$ (2 - 10\,keV).
Same data as in Fig.~\ref{LRLBAT} only converted between 14 - 195\,keV and 2 - 10\,keV bands using Eq.~\ref{BATtoX}.
Data include the sources from Tables \ref{flux} (black), and \ref{previous} (red). 
Upper limits are marked with green down-pointing arrows. 
The $L_{\rm mm}  =  10^{-4}L_{\rm X}$ line is plotted for reference.
}
\label{LRLX}
\end{figure}

In order to compare directly with the 2--10~keV band, we converted all $L_{\rm BAT}$ luminosities to $L_{\rm X}$ using Eq.~\ref{BATtoX}.
The $L_{\rm mm} $ vs. $L_{\rm X}$ values are plotted in Fig.~\ref{LRLX}.
Here too, we plotted the $L_{\rm mm}  = 10^{-4}L_{\rm X}$ line to guide the eye.
As expected, $L_{\rm mm} $ also approximately correlates with $L_{\rm X}$.
As opposed to $L_{\rm BAT}$ and Fig.~\ref{LRLBAT}, the $L_{\rm X}$ data points lie somewhat above the line. 
Indeed, fitting the normalization in Fig.~\ref{LRLX} yields $L_{\rm mm}  = 10^{-3.72\pm 0.02}L_{\rm X}$, 
but the small sample and the scatter impede a strong conclusion.
One of the possible reasons for the scatter in mm vs X-ray luminosities may be the lack of broad spectral coverage, without which $\nu L_\nu$ is only a poor estimate of the true mm luminosity.

A fit to the slope in Fig.~\ref{LRLX} yields $0.86\pm 0.04$, to be compared with $1.1 \pm\ 0.15$ for 1.4\,GHz in \citet{panessa15}.
The different slopes of the 1.4~GHz and the 100~GHz luminosities, with respect to the X-ray luminosity could suggest that the cm- and mm- wave sources represent different components, from different size scales of the AGN.
This is supported by the high-frequency excess that suggests a separate mm-wave AGN component.

Interestingly, the presumably RL sources in the present sample (see Sec.~\ref{Introduction}), do not show evidence of radio loudness in their $L_{\rm mm}  / L_{\rm X}$ values (Fig.~\ref{LRLX}). 
We suspect this is a result of a compact radio/mm core that is independent of more extended radio emission from a jet.
The one outlier is the suspected LINER NGC\,5106 at the very top of Figs.~\ref{LRLBAT},~\ref{LRLX}.
NGC\,5106 is radio loud  in terms of its 5\,GHz to B-band flux density ratio, and 
its value of $\log (L_{\rm mm}  / L_{\rm X}) =  -2.37$ is high compared to the present X-ray selected sample,
though not unusual for low-luminosity AGN  \citep[c.f.,][]{doi11, Wu18}.
\citet{Parisi14} point to contradicting indicators regarding the classification of NGC\,5106 (PBC J1321.1+0858 there).
Its optical line ratios put it in the region between Seyferts, LINERs, and star forming galaxies.
Its [O\,I] optical line is relatively weak for a Seyfert.
Indeed, NGC\,5106 has a 20\arcsec extended UV nucleus, and its FIR emission indicates an unusually high star formation rate of above 20 solar masses per year, one of the three highest of the \bat\ AGN sample \citep{Melendez14}. 
We did not find any high-resolution radio images of NGC\,5106, but its high $L_{\rm mm}  / L_{\rm X}$ ratio could be explained if it has a compact radio jet.


\section{Discussion \& prospects}
\label{sec:discussion}

We observed 26 hard X-ray bright AGN at 100\,GHz with \carma\ and detected 18 mostly unresolved cores.
This sample increases the number of RQ AGNs observed in mm waves by a factor of a few.
Most of the sources feature a 100\,GHz excess with respect to the extrapolation of the low-frequency steep spectral slope. 
This spectral turnover adds to previous evidence of an optically thick mm-wave AGN component that is superimposed on the steep-slope, optically thin, likely more extended component at lower radio frequencies. 

If the mm-wave emission mechanism is synchrotron self-absorption, the deduced sizes $R_{\rm pc}$ of an optically thick 100\,GHz synchrotron source at the observed luminosities, are in the range of $10^{-4} - 10^{-3}$~pc.
This size estimate is of the order of the typical X-ray variability time scale of approximately a light day, which suggests perhaps that the mm-wave component is associated with the X-ray source.

\begin{figure}
\vskip -1cm
\hskip -1.4cm
\includegraphics[scale=0.38,angle=180, angle=180]{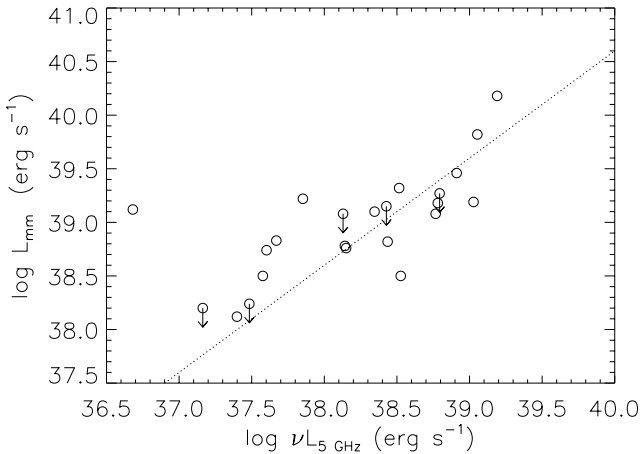} 
\vskip -0.7cm
\caption{Relation between $L_{\rm 5GHz}$ and $L_{\rm 100GHz}$ luminosities for 23 targets from Tables \ref{flux}, \ref{previous} that have a \vla\ 5\,GHz measurement in the A configuration. Upper limits are indicated with arrows. The dotted line is the best fit, indicating a broad-band spectral slope of --0.6. The outlier on the left is the radio-loud NGC\,2110.}
\label{L5_L100}
\end{figure}

Furthermore, the 100\,GHz luminosity appears to be roughly correlated with the X-ray luminosity, both in the \bat\ band of 14--195~keV, and with the implied 2--10~keV luminosity. 
The measured luminosities follow what is an $L_{\rm mm}  =  10^{-4}L_{\rm X}$ relation, which is not yet statistically significant. 
At 22\,GHz, \citet{smith16} found  $L_{\rm R}  =  10^{-5} - 10^{-4} L_{\rm X}$, which seems like an extension of the $L_{\rm R}  =  10^{-5} L_{\rm X}$ correlation found for the 5\,GHz luminosity.
One may wonder if $L_{\rm mm}$ also correlates with $L_{\rm R}$ at 5\,GHz directly. For this we plot the two luminosities in Fig.~\ref{L5_L100}, for those sources that have a \vla\ measurement at 5\,GHz with the A-array. An overall correspondence between the two bands can be seen. A broad-band spectral slope of $\alpha = 0.6$ is implied.
This is expected given the generally steep spectra of the sample (see Table~\ref{flux}), including many even steeper slopes, hence the mm-wave excess.
The correlation with X-rays is interestingly similar to the correlation found for coronally active stars, and suggests that the X-ray and radio sources may be both associated with stochastic magnetic activity.

Other interpretations of the mm-wave excess are also possible, and more spectral and temporal information is still needed to discern between the various scenarios. A superposition of many compact galactic sources, e.g. supernova remnants (SNRs), is unlikely, because the 1.4\,GHz luminosity of an SNR is $\sim 10 ^{30 - 32}$\,erg\,s$^{-1}$ \citep{Pannuti07, Lee14}, and less at 100\,GHz, requiring millions of them to account for the observed radio and mm-wave luminosity. We note also that the  $L_{\rm R} / L_{\rm X}$ ratio of SNRs implied from these papers is closer to 10$^{-6}$, so different from AGN. X-ray binaries have even lower $L_{\rm R} / L_{\rm X}$ ratios. The most extreme lowest luminosity AGNs do start approaching luminosities of SNRs and X-ray binaries \citep{Wu18}, but not the luminous \bat\ AGNs considered here. Furthermore, if the spectrum from radio to mm is flat \citep[c.f.][]{Wu18} it could be due to free-free emission ($F_\nu \propto \nu ^{-0.1}$) from the AGN, e.g., the BLR. 
A free-free absorbed weak jet, analogous to RL AGN, would then also be possible to account for the mm-wave emission \citep{doi16}, but the correlation with the X-ray luminosity would require a different explanation. In the present sample, except for NGC\,2110, which is radio loud and has extended radio emission, strong jet scenarios are ruled out.
Due to the lack of sufficient spectral information, another intriguing source of mm-wave emission might be charged spinning dust grains whose rotational dipole emission peaks at $\sim 30$\,GHz, and is much weaker at few GHz \citep{Draine98}.
These have been invoked to explain anomalous microwave emission in our galaxy \citep{Leitch97}, a weak excess over the cosmic microwave background. 
A survey of the 10 - 100\,GHz spectra would be needed to assess their contribution, but here too, the relation to X-rays would require a separate explanation.

Despite the mounting circumstantial evidence, direct measurements of 100\,GHz variability, and its relation to the X-ray variability, are still required if one is to test the AGN connection between the mm band and the X-ray sources, namely the accretion disk coronae.
In order to decisively test the conjecture of radio/mm emission from the accretion disk corona, 
simultaneous mm (not few-GHz radio that hardly varies) and X-ray monitoring is needed.
Temporal correlation between the mm and X-ray light curves would be the smoking gun of coronal radio/mm emission.
An example of such a correlation in stellar coronae is the Neupert effect \citep{neupert68, guedel02}, 
by which the time derivative of the X-ray light curve correlates with the radio light curve.
This effect is explained by non-thermal, radio electrons spiraling down the stellar (would be accretion disk in AGN) corona, and their time-integrated energy heating the stellar chromosphere (accretion disk) to produce X-rays.  

Finally, a systematic study of broadband radio and mm spectra of a full sample of RQ AGNs between 1--100\,GHz is desired, in order to better characterize the high-frequency excess and its various manifestations in different AGN.
The present work as well as those of \citet{smith16}, \citet{doi16}, and an ongoing 45\,GHz campaign on the PG~quasars (Baldi et~al. in preparation), should all contribute to the large picture of the physical processes behind radio and mm-wave emission in RQ AGNs.

\section*{Acknowledgments}

The research at the Technion is supported by the I-CORE program of the Planning and Budgeting Committee (grant number 1937/12) and by a grant from the Asher Space Research Institute. 
EB is grateful for the warm hospitality and support during a sabbatical visit to the University of Maryland, College Park, and for funding from the European Union's Horizon 2020 research and innovation programme under the Marie Sklodowska-Curie grant agreement no. 655324. 
EB thanks Francesca Panessa for important comments on the manuscript. 
This work was performed in part at the Aspen Center for Physics, which is supported by National Science Foundation grant PHY-1607611 and by a grant from the Simons Foundation. 

\label{lastpage}

\bibliography{my}

\end{document}